\documentclass[aps,prl,superscriptaddress,twocolumn]{revtex4-1}

\usepackage{graphicx}
\usepackage{fancyhdr}
\usepackage{dcolumn}
\usepackage{bm}
\usepackage{hyperref}
\usepackage[ 
margin=0.75in,
]{geometry}

\graphicspath{{.}}

\usepackage[usenames]{color}

\DeclareMathAlphabet{\mathpzc}{OT1}{pzc}{m}{it}

\def\sg{Fm$\overline 3$m}
\begin{document}

	\title{Emphanitic anharmonicity in {PbSe} at high temperature and the anomalous electronic properties in the {PbQ} ({Q=S}, {Se}, {Te}) system}

	\author{Runze Yu}
    \affiliation{Condensed Matter Physics and Materials Science Department, Brookhaven National Laboratory, Upton, NY 11973, USA}
	
	\author{Emil S. Bozin}
	\email{bozin@bnl.gov}
	\affiliation{Condensed Matter Physics and Materials Science Department, Brookhaven National Laboratory, Upton, NY 11973, USA}
	
	\author{Milinda Abeykoon}
	\affiliation{Photon Sciences Division, Brookhaven National Laboratory, Upton, NY 11973, USA}

	\author{Boris Sangiorgio}
	\author{Nicola A. Spaldin}
	\affiliation{Materials Theory, ETH Zurich, Wolfgang-Pauli-Strasse 27, CH-8093 Z\"urich, Switzerland}

	\author{Christos D. Malliakas}
	\affiliation{Department of Chemistry, Northwestern University, Evanston, IL 60208, USA}

	\author{Mercouri G. Kanatzidis}
	\affiliation{Department of Chemistry, Northwestern University, Evanston, IL 60208, USA}
	\affiliation{Materials Science Division, Argonne National Laboratory, Argonne, IL 60439, USA}

	\author{Simon J. L. Billinge}
	\affiliation{Condensed Matter Physics and Materials Science Department, Brookhaven National Laboratory, Upton, NY 11973, USA}
	\affiliation{Department of Applied Physics and Applied Mathematics, Columbia University, New York, NY 10027, USA}
	
	\date{\today}

	\begin{abstract}
	The temperature dependence of the local structure of PbSe has been investigated using pair distribution function (PDF) analysis of
 x-ray and neutron powder diffraction data and density functional theory (DFT) calculations. Observation of non-Gaussian PDF peaks at high temperature indicates the presence of significant anharmonicity, which can be modeled as Pb off-centering along [100] directions that grows on warming similar to the behavior seen in PbTe and PbS and sometimes called emphanisis. Interestingly, the emphanitic response is smaller in PbSe than in both PbS and PbTe indicating a non-monotonic response with chalcogen atomic number in the PbQ (Q=S, Se, Te) series. The DFT calculations indicate a correlation between band gap and the amplitude of [100] dipolar distortion, suggesting that emphanisis may be behind the anomalous composition and temperature dependencies of the band gaps in this series.		
	\end{abstract}

	\maketitle

PbQ (Q=S, Se, Te) is an important thermoelectric system~\cite{dugha;pb02}. Also notable about this system is an anomalous temperature and composition dependence of the electronic band gap, i.e., the energy gap increases with increasing temperature for all three members of the PbQ series,\cite{gibbs_temperature_2013,cardona_isotope_2005,baleva_temperature_1990} and shows a non-monotonicity with chalcogen atomic number~\cite{taube;jap66,dalven_energy-gap_1971,pei_electrical_2012,nie_band_2016}, neither of which behaviors are observed in other binary compound semiconductors~\cite{long;b;ebis68,tsay;jpcs73,bauma;pssb74}. PbTe has also recently garnered research interest because of the observation of appreciable Pb anharmonicity at high temperatures, sometimes called emphanisis, that results in anomalously large excursions away from the high-symmetry average positions of the rock-salt structure~\cite{bozin;s10,Sangiorgio2017,ghezz;pssb73,subha;pram78,lebed;poss99,zhang;prl11}.
While seen in other lone-pair materials~\cite{knox;prb14,fabin;jacs16} and therefore presumably related to the stereochemical activity of the Pb$^{2+}$ lone pair, the precise nature and origins of this effect are believed to be strongly associated to the presence of the 6s$^2$ lone pair in Pb$^{2+}$. Here, we extend the study of the PbQ series to explore emphanisis in PbSe, and use the results to explore the relationship between the dynamic Pb off-centering and the band gap in the PbQ system.

There are increased amplitude atomic motions in all materials with increasing temperature, but what is remarkable in emphanitic systems is the large amplitude of the fluctuations, as large as 0.25~\AA~\cite{bozin;s10}, and the extreme anharmonicity~\cite{delai;nm11,jense;prb12}. On average, the fluctuations do not break the long range symmetry, as evident by the preservation of the average crystallographic cubic structure and the failure to see a net off-centering on average in EXAFS~\cite{keibe;prl13}. Inelastic neutron scattering (INS) measurements~\cite{delai;nm11,jense;prb12} showed that the atomic displacements are dynamic and that there is significant anharmonicity in the dynamics, consistent with the non-Gaussian atomic pair distribution function (PDF) peaks. For example, Delaire~{\it et al.}~\cite{delai;nm11} showed the appearance of an avoided crossing behavior in the phonon dispersions, as well as an anomalous lowering and damping of the longitudinal acoustic phonons and a “waterfall” effect at the zone center, consistent with strong anharmonicity, and Jensen~{\it et al.}~\cite{jense;prb12} identified the appearance on warming of a new dynamic mode at $\sim 6$~meV that suggested a dynamic symmetry breaking. First principles calculations~\cite{bozin;s10,delai;nm11,shiga;prb12} also indicate the presence of strong anharmonic effects in these materials, and, in combination with diffuse scattering measurements,\cite{Sangiorgio2017} have recently clarified the existence and nature of correlated local dipolar ordering in PbTe.

Emphanisis was originally reported in PbTe and PbS. Here we complete the investigation of the PbQ (Q = S, Se, Te) series by reporting results from the PbSe system, and comparing the behavior across the series. We carried out complementary x-ray and neutron pair distribution function analysis. We show that PbSe has a response very similar to PbS and PbTe and is also emphanitic. Interestingly, in PbSe the refined amplitude of the dynamic displacements at high temperature is smaller and the PDF remains harmonic to higher temperature than in either PbS or PbTe.  indicating that the strength of the emphanisis across the series is non-monotonic with chalcogen atomic number, being weaker in PbSe than in either PbS or PbTe. This may explain the anomalous non-monotonicity of the band gap in this series of materials~\cite{taube;jap66,dalven_energy-gap_1971,pei_electrical_2012,nie_band_2016}, though the underlying origin of the non-monotonicity of the emphanisis is not clear.

The PbSe, PbS, and PbTe samples were prepared by methods previously reported~\cite{bozin;s10}. The resulting polycrystalline samples were pulverized for total scattering experiments. The experiments were performed at the NPDF beamline at the Lujan Center at Los Alamos National Laboratory and the 28-ID-2 beam line of the National Synchrotron Light Source-II (NSLS-II) at Brookhaven National Laboratory (BNL). Data were collected over wide temperature ranges, $15\le T\le 550$~K and $10\le T\le 480$~K for neutron and x-ray total scattering, respectively. The neutron and x-ray data reduction to obtain the PDFs was carried out using the PDFgetN~\cite{peter;jac00} with $Q_{max} = 28$~\AA\ and xPDFsuite software with $Q_{max} = 30$~\AA,  respectively~\cite{yang;arxiv15,egami;b;utbp12}, using standard methods.~\cite{egami;b;utbp12,juhas;jac13} The data were modeled using PDFgui~\cite{farro;jpcm07} with a cubic rocksalt structure model (space group \sg).
Our first-principles calculations were performed using the PAW~\cite{bloechl1994,kresse1999} implementation of density functional theory (DFT) as in the VASP package~\cite{kresse1996} (for further details see SI).

\begin{figure}
\includegraphics[width=\columnwidth]{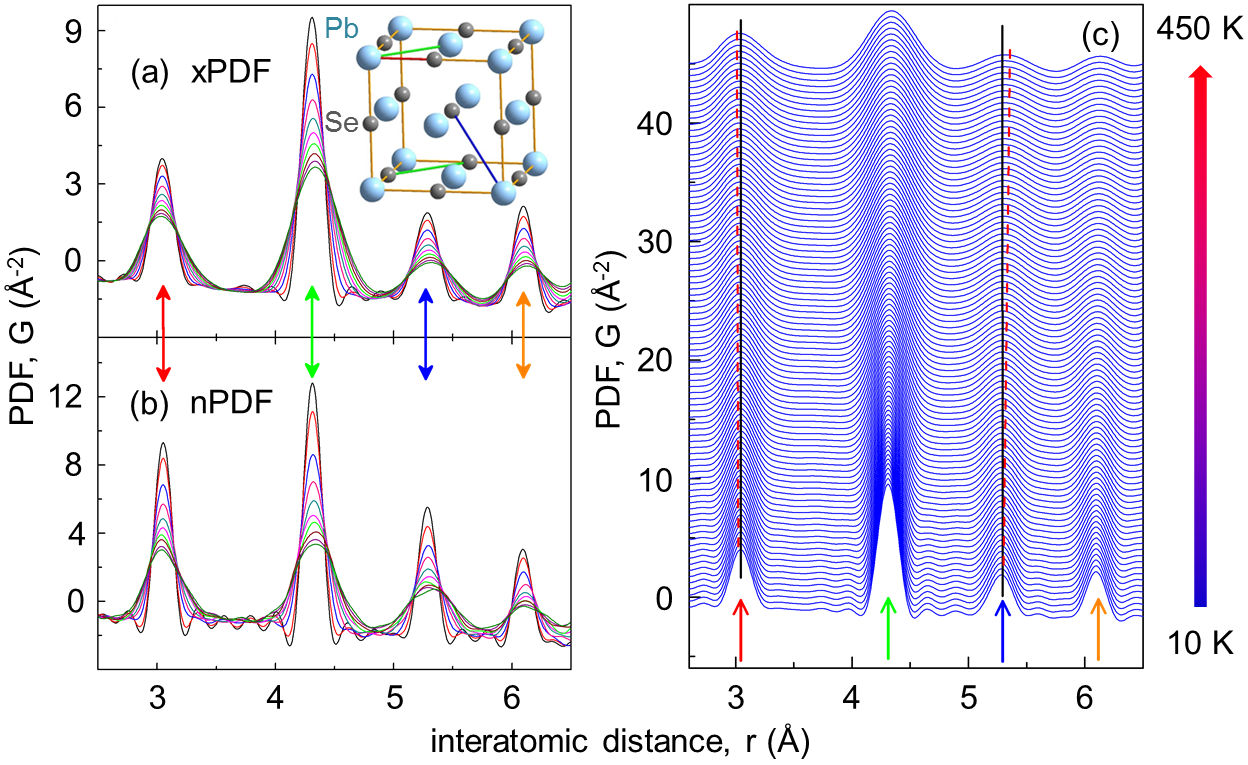}
\caption{Temperature evolution of PbSe PDF patterns over 10~K -- 450~K range in 50~K increments obtained by (a) x-ray (xPDF) and (b) neutron (nPDF) total scattering. Inset to (a) shows \sg\ structure of PbSe with interatomic distances color coded. (c) Waterfall representation of x-ray $G(r)$. Data are offset for clarity. Vertical solid black lines mark the PDF peak positions at base temperature. Sloping dashed red lines track the apparent PDF peak centroids with temperature.}
\label{fig:PDFs}
\end{figure}

The emphanitic effects are substantial and it is possible to see the appearance of non-Gaussian PDF line-shapes and anomalous temperature-dependent peak shifts by eye~\cite{bozin;s10,knox;prb14} as evident in the low-$r$ region of PbSe PDF experimental data from x-ray and neutron measurements
shown in Figure~\ref{fig:PDFs}.
%
%
All the peaks significantly broaden with increasing temperature and become highly non-Gaussian at higher temperatures, similar to PbTe and PbS~\cite{bozin;s10}. The nearest-neighbor peak becomes asymmetric and drops as rapidly in height as the higher-neighbor peaks, an effect that is unusual in PDFs since correlated motion effects tend to sharpen the nearest neighbor correlations with respect to the others~\cite{jeong;jpc99}. In addition, the third peak also shifts anomalously to higher $r$ (Fig.~\ref{fig:PDFs}(c)). All these observations are characteristic of emphanisis.

\begin{figure}
\includegraphics[width=\columnwidth]{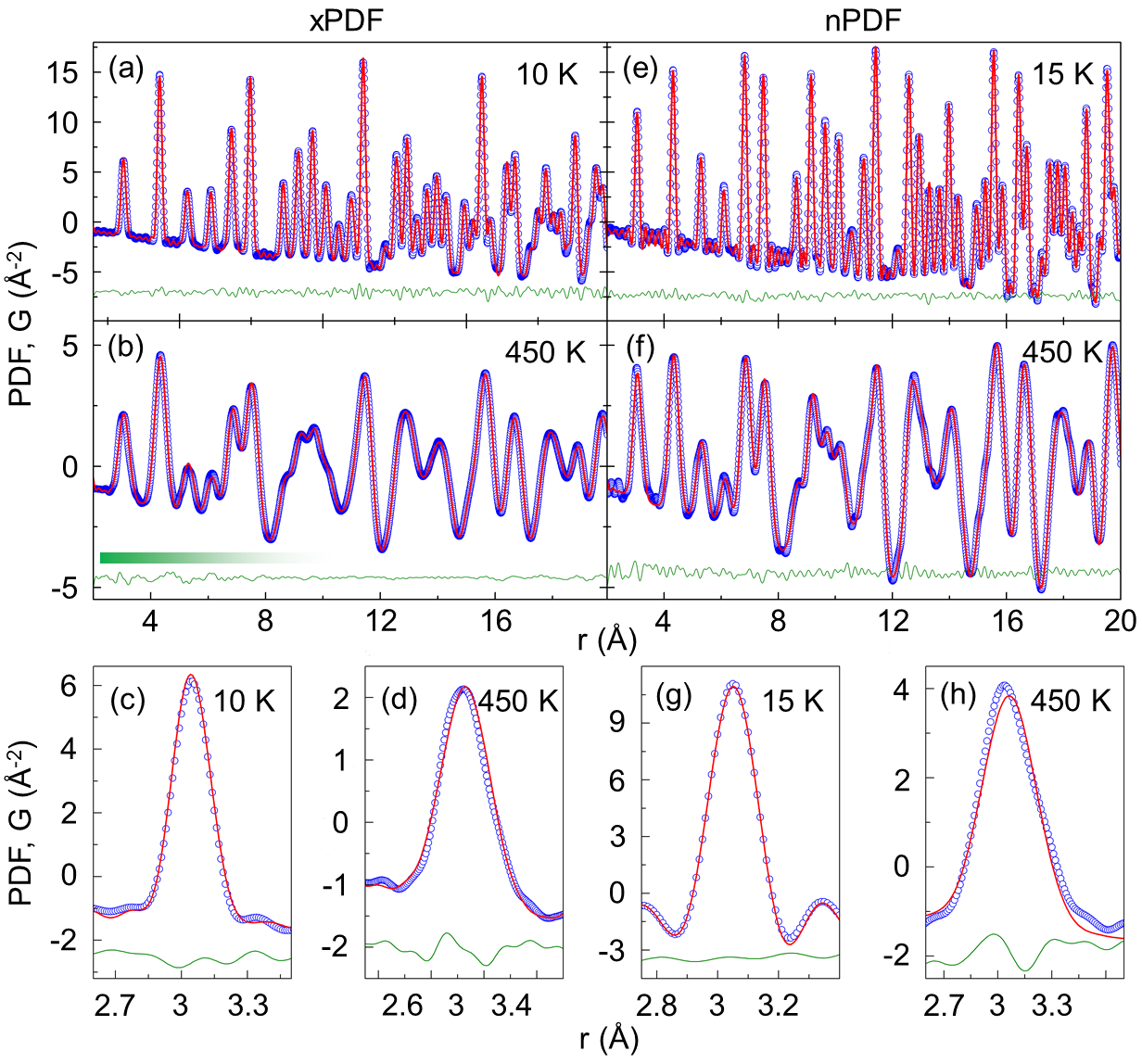}
\caption{Fits of \sg\ model (solid red line) to experimental PDFs (open blue symbols) obtained by (a)-(d) x-ray probe and (e)-(h) neutron probe. Top panels show broad r-range view, bottom panels focus on nearest neighbor distributions. Solid green lines are the differences (offset for clarity). Temperature is as indicated. Shaded green rectangle in (b) sketches crossover from the local to average behavior.}
\label{fig:modeling}
\end{figure}

To investigate this behavior more quantitatively the xPDF and nPDF data were fit with the rocksalt structure model, Figure~\ref{fig:modeling}. At low temperature the cubic model explains the data well at all length scales (xPDF, Fig.~\ref{fig:modeling}(a) and nPDF Fig.~\ref{fig:modeling}(e)), confirming that there are no detectable distortions at this temperature.
%
%
At high temperature the \sg\ model explains the average structure well, and is also consistent with the data on intermediate length scales. The model is less successful on short length scales, for example, visible below 6~\AA\ as increased amplitude features in the green difference curve. Fits to the 450~K data are shown in Fig.~\ref{fig:modeling}(b) and~\ref{fig:modeling}(f). They indicate the presence of significant local distortions on a length scale of a few unit cells, where we note that the PDF cannot, by itself, determine if these distortions are static or dynamic. Over the long range the locally distorted structure averages to the rock-salt structure. In Fig.~\ref{fig:modeling}(c), (d), (g), and (h), we show the Pb-Se nearest neighbor PDF peaks on an expanded scale. At low temperature, Fig.~\ref{fig:modeling}(c) and~\ref{fig:modeling}(g), the peaks appear as sharp, well defined single-Gaussian functions with small termination ripples on each side originating from the finite $Q$ range of the Fourier transform~\cite{egami;b;utbp12}. The red lines are calculated PDF profiles based on the rock-salt model having a pure Gaussian line-shape, convoluted with a sinc function to account for the termination effects. This is characteristic of a single average bond length with harmonic motion taking place around that position, indicating that the ground state of PbSe at low temperature is the expected ideal rock salt in both the local and average structures. However, at 450~K (Fig.~\ref{fig:modeling}(d) and~\ref{fig:modeling}(h)), the peaks are considerably broadened and qualitatively non-Gaussian, with intensity shifted to the high-$r$ side of the peak. These observations unambiguously point to the appearance of significant anharmonic effects in PbSe with increasing temperature, and demonstrate that emphanitic behavior is universally seen in all three lead chalcogenides, PbS, PbSe, and PbTe.

\begin{figure}
\includegraphics[width=\columnwidth]{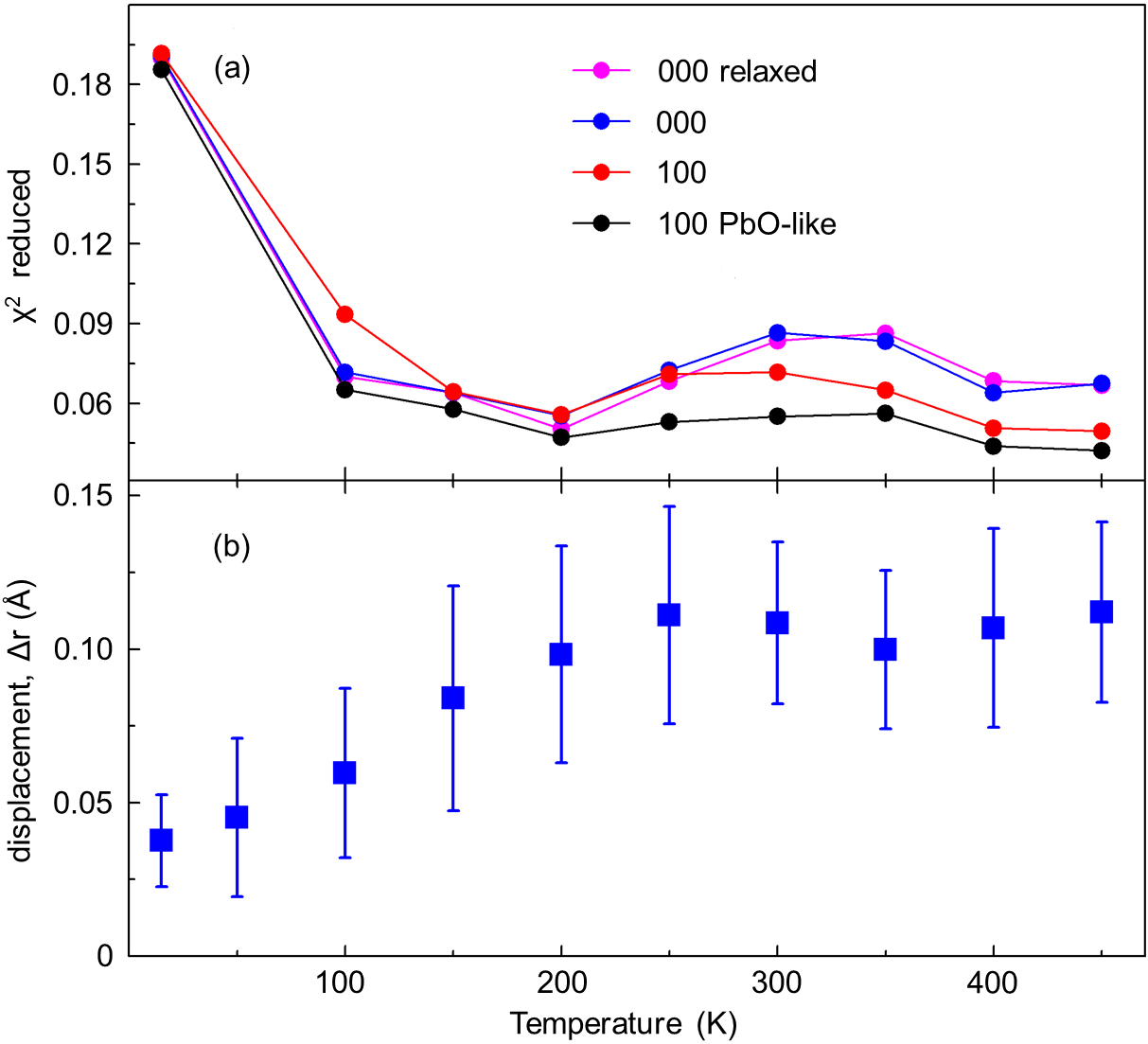}
\caption{(a) Assessment of different models (as indicated) for the local structure at various temperatures as seen by reduced $\chi^{2}$ of the fits. (b) Temperature evolution of estimated Pb local off-centering amplitude obtained from model with 100 PbO-like displacements. Fits were done to neutron PDF data and utilized the same protocols as those in the original report on PbTe and PbS~\cite{bozin;s10}, to allow direct comparison.}
\label{fig:distortions}
\end{figure}

Following the same approach as was taken in the initial study~\cite{bozin;s10}, and in order to quantify the underlying bondlength distribution, the local structure was further explored by fitting experimental PDFs with several undistorted and distorted models, as described in the supplement. The cubic model with Pb constrained to remain on its crystallographic positions (undistorted-000) and a model where the Pb is not allowed to displace off its high symmetry positions but the unit cell may take on a tetragonal distortion, as well as models where the Pb ion can display off its high symmetry position in different directions ([110] or [100] displacements) were all tested~\cite{bozin;s10}. Just as observed in PbTe and PbS, we find that a model allowing displacements of Pb along [100] directions, similar to the PbO structure, is preferred at high temperature, consistent with the atomic probability distribution being highly non-Gaussian and appreciably elongated along the [100] direction (Fig.~\ref{fig:distortions}(a)), i.e., in the emphanitic state at high temperature the Pb ions spend significant amount of time away from the high-symmetry central position in the form of fluctuating dipoles.
%
%

We note that this type of modelling does not imply static displacements for Pb.  The PDF yields the instantaneous structure and any offset may be static or dynamic, or have contributions from both.  There is ample evidence in PbQ that these displacements are dynamic in nature.  It also does not necessarily imply that the time-average of the displacements is off-centered.  In other words the {\it time or ensemble average} atomic probability distribution may be peaked at the center.  It does imply that the Pb ions are making large excursions from the average position, and spending significant amount of time away from the central position.  Whether or not these excursions result in local dipoles depends on how the excursions are correlated between neighboring sites.  If the excursions are correlated between neighboring sites, it implies the formation of local (and in general fluctuating) polar nanoregions.  If the excursions are anti-correlated the material would be locally anti-ferrodistortive.  Our 1D PDF data are not sufficiently sensitive to detect the presence or nature of correlations between neighboring displacements with any certainty, but a recent 3D $\Delta$-PDF measurement does indeed detect that neighboring displacements tend to be correlated~\cite{Sangiorgio2017}.

We extracted the amplitude of the local off-centering in the [100] direction that is needed to reproduce the data within this PbO-like model. Following the procedure carried out for PbTe and PbSe~\cite{bozin;s10}, we separate ``normal'' (harmonic) dynamics from the emphanitic dynamics by fitting a Debye curve~\cite{debye;adp12} to the low temperature part of the atomic displacement parameter (ADP) data, below 200~K, and extrapolating it to high temperature. The Pb isotropic ADPs in the model are then fixed to the extrapolated Debye value, and any additional distortion that we may ascribe to the emphanisis is accounted for in the Pb [100] off-centering. This is justified as a way to include in a highly constrained small-box model the non-Gaussian anharmonic behavior superposed on top of any increased harmonic displacement amplitudes of the structure. Doing the modeling in this fashion also allows us to directly compare the results to earlier work on the PbTe and PbS which were analyzed this way. The refined Pb off-centering increases from zero at low-temperature to a value of about 0.12~\AA\ at high temperature (Fig.~\ref{fig:distortions}(b)). This is a large amplitude distortion, but significantly, is the smallest distortion among the PbQ series of compounds (PbS = 0.25~\AA , PbTe = 0.24~\AA)~\cite{bozin;s10}. This implies that there is a non-monotonicity of the emphanisis with chalcogen atomic number on going from PbS, through PbSe to PbTe, with the PbSe being the leasat emphanitic of the three.

To further explore the non-monotonicity of the emphanisis across the series PbS-PbSe-PbTe we consider other measures. The anharmonicity in the motions of the Pb atoms may be seen as an anomalous non-thermal increase in the ADP of Pb at high temperature.  It was noticed~\cite{bozin;s10} that this temperature dependence is consistent with the ADPs following a Debye model behavior at low temperature, and following the same Debye model at high temperature, but with the Debye curve offset upwards to explain the high-T data.  The offset parameter, $\Delta$U$_{off}$, is taken as a measure of the additional ``non-thermal" displacement amplitude coming from the anharmonicity.   $\Delta r_{RMS}=\sqrt{\Delta U_{off}}$ from each of the three compounds is displayed in Fig~\ref{fig:comp}(d), (see supplementary information (SI) for details).
%
\begin{figure}
\includegraphics[width=\columnwidth]{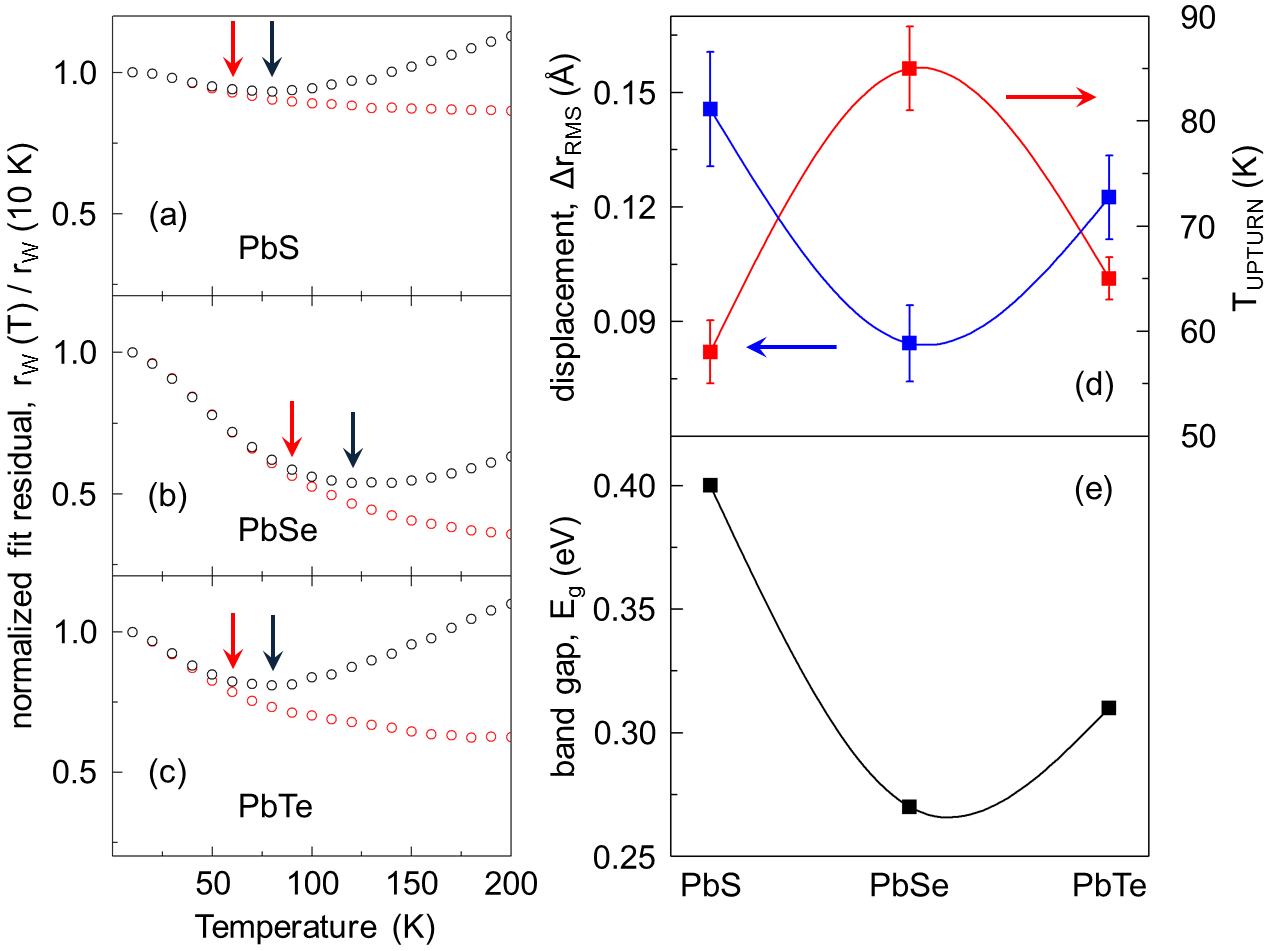}
\caption{(a)-(c) Normalized fit residual, $r_{W}$, of \sg\ model fit to xPDF data of PbS, PbSe, and PbTe (as indicated) over a range of temperatures. Open red symbols are obtained from fits carried over $r$-range of PDF sensitive solely to average behavior (10--50~\AA), excluding the short $r$-range. Open black symbols originate from fits where short range local structural information in PDF was included in the fitting. Vertical red arrows mark upturn temperature at which the two trends separate, and vertical black arrows mark temperature of the apparent minimum in the short-range sensitive trends. (d) Root mean square displacement in PbS, PbSe, and PbTe, estimated from the differential static offset of the Debye model as this is fit to the temperature dependence of the Pb isotropic atomic displacement parameter (left ordinate). Temperature of the apparent upturn in normalized fit residuals shown in (a)-(c) (right ordinate). See text for details. (e) Reported energy gaps of lead chalcogenides PbS, PbSe, and PbTe~\cite{androulakis_thermoelectrics_2011,androulakis_high-temperature_2011,wang_heavily_2011,johnsen_nanostructures_2011,pei_electrical_2012}.}
\label{fig:comp}
\end{figure}
%
Among the three chalcogen series, PbSe has the smallest $\Delta$U$_{off}$, and therefore the smallest distortion, in agreement with the direct refinements of displacements in the PbO distorted models. Another way to quantify the strength of the emphanitic effects in each compound is by the temperature where the anharmonicity first becomes evident in the structure. One measure of this is the temperature at which fits to the PDF data of the undistorted cubic model first become inadequate as temperature increases, as measured by a goodness of fit parameter such as weighted fit residual $R_w$. As temperature increases from 10~K the $R_w$ of the cubic model initially decreases. This decrease in $R_w$ on warming is commonly observed in materials when the structural model correctly describes the structure. A good fit is obtained at all temperatures, but the $R_w$ decreases somewhat on warming because the PDF peaks broaden and therefore become easier to fit. This $R_w$ lowering effect normally continues to the highest temperature, as indeed is seen to be the case in our data for the \sg\ model fit over long length scale, but excluding the short range data, shown in red in Fig.~\ref{fig:comp}(a)-(c). However, when fitting the low-$r$ region, the $R_w$ of the cubic model goes through a minimum and then starts to increase.  This minimum in $R_w$ gives a characteristic temperature at which the harmonic cubic model is becoming inadequate to explain the local structure. The blue and red arrows in the figure indicate these points for each of the chalcogens, and it is clear that in PbSe the sample has to warm to a higher temperature than PbS and PbTe before significant anharmonicity is observed, suggesting that the anharmonic effects are less in that system. This crossover temperature, T$_{upturn}$, is shown in red in Fig.~\ref{fig:comp}(d) vs. chalcogen atomic number. The chalcogen dependence of the band gap for the series extracted from the literature~\cite{androulakis_thermoelectrics_2011,androulakis_high-temperature_2011,wang_heavily_2011,johnsen_nanostructures_2011,pei_electrical_2012} is shown in Fig.~\ref{fig:comp}(e).

It is tempting to speculate on a possible relationship between the non-monotonicity of the band gap and the strength of the emphanisis. To explore this we used density functional theory (DFT) to compute the dependence of the band gap on Pb [100]-displacement amplitude in the lead chalcogenides. Indeed, the calculations indicate that the computed band gap increases with increasing Pb off-centering (see SI).
We compared this evolution with the behavior of rocksalt-structure NaCl, as well as a series of perovskite-structure materials (PbTiO$_3$, BaTiO$_3$, and LaAlO$_3$). For both structure types we observe a larger increase of the band gap in the systems possessing a lone pair. In the rocksalt materials, emphanitic displacements in  PbQ have a larger effect than in NaCl and in the perovksites the band-gap effect is larger in PbTiO$_3$ than in BaTiO$_3$ (a ferroelectric without lone pair) and very small in LaAlO$_3$ which shows no tendency for a ferroelectric distortion.
Thus, the stereochemical activity of the lone pair, be it static (PbTiO$_3$) or dynamic (PbQ)
can be seen to correlate with the increase of the band gap.
We propose that this link between the emphanitic effects and the band gap could serve to explain both the poorly understood non-monotonic chalcogen dependence of $E_g$ in this series, but also the anomalous temperature dependence of the band gap. The emphanitic effects therefore have a significant effect on the electronic properties of the materials, as well as through increased scattering.   The band gaps of all these materials increase with increasing temperature, which is the opposite of the behavior expected for semiconductors~\cite{dey_origin_2013}, but would be well explained by the increase in the emphanitic effects with increasing temperature and the positive correlation of band gap to those effects that is evident in the DFT calculations.  A similar emphanitic effect associated with the 5s$^2$ lone pair on Sn$^{2+}$ in the perovskite halide CsSnBr$_3$ was reported recently and in this case too the widening of the energy gap $E_g$ with rising temperature was linked to the increasing off-center displacement of the Sn$^{2+}$ atom~\cite{fabin;jacs16}.

	\section*{Acknowledgments}
Work at Brookhaven National Laboratory was supported by U.S. Department of Energy, Office of Science, Office of Basic Energy
Sciences (DOE-BES) under contract DE-SC00112704. Work at Argonne National Laboratory was supported by the U.S. Department of Energy, Office of Science, Materials Sciences and Engineering.
The neutron diffraction measurements were carried out at NPDF instrument of the Lujan Neutron Scattering Center at Los Alamos National Laboratory, and the x-ray experiments were carried out at beamline 28-ID-2 (XPD) of the National Synchrotron Light Source II at Brookhaven National Laboratory. Use of the National Synchrotron Light Source II, Brookhaven National Laboratory, was supported by DOE-BES under contract No. DE-SC0012704. BS and NAS acknowledge support from ETH Z\"urich, the ERC Advanced Grant program (No. 291151), and the Swiss National Supercomputing Centre (CSCS) under project ID s624.



\end{document}